\documentclass[lettersize,journal]{IEEEtran}
\usepackage{cite}
\usepackage{amsmath,amssymb,amsfonts}
\usepackage{algorithmic}
\usepackage{graphicx,color}
\usepackage{textcomp}
\usepackage{amsmath,amsfonts}
\usepackage{algorithmic}
\usepackage{algorithm}
\usepackage{array}
\usepackage[caption=false,font=normalsize,labelfont=sf,textfont=sf]{subfig}
\usepackage{textcomp}
\usepackage{stfloats}
\usepackage{url}
\usepackage{verbatim}
\usepackage{graphicx}
\usepackage{cite}
\usepackage{hyperref}
\usepackage{subfig}
\begin{document}

\title{Embracing Radiance Field Rendering in 6G: Over-the-Air Training and Inference with 3D Contents}

\author{Guanlin Wu, Zhonghao Lyu, Juyong Zhang, and Jie Xu
\thanks{Part of this paper has been submitted to IEEE International Symposium on Wireless Communication Systems (ISWCS) 2024 \cite{guanlin}.} 
\thanks{G. Wu, Z. Lyu, and J. Xu are with the School of Science and Engineering (SSE) and the Future Network of Intelligence Institute (FNii), The Chinese University of Hong Kong (Shenzhen), Guangdong 518172, China (e-mail: guanlinwu1@link.cuhk.edu.cn, zhonghaolyu@link.cuhk.edu.cn, xujie@cuhk.edu.cn). J. Xu is the corresponding author.}
\thanks{J. Zhang is with the School of Mathematical Science, University of Science and Technology of China, Hefei 230026, China (e-mail: juyong@ustc.edu.cn).}}

\maketitle

\begin{abstract}
The efficient representation, transmission, and reconstruction of three-dimensional (3D) contents are becoming increasingly important for sixth-generation (6G) networks that aim to merge virtual and physical worlds for offering immersive communication experiences. Neural radiance field (NeRF) and 3D Gaussian splatting (3D-GS) have recently emerged as two promising 3D representation techniques based on radiance field rendering, which are able to provide photorealistic rendering results for complex scenes. Therefore, embracing NeRF and 3D-GS in 6G networks is envisioned to be a prominent solution to support emerging 3D applications with enhanced quality of experience. This paper provides a comprehensive overview on the integration of NeRF and 3D-GS in 6G. First, we review the basics of the radiance field rendering techniques, and highlight their applications and implementation challenges over wireless networks. Next, we consider the over-the-air training of NeRF and 3D-GS models over wireless networks by presenting various learning techniques. We particularly focus on the federated learning design over a hierarchical device-edge-cloud architecture, which is suitable for exploiting distributed data and computing resources over 6G networks to train large models representing large-scale scenes. Then, we consider the over-the-air rendering of NeRF and 3D-GS models at wireless network edge. We present three practical rendering architectures, namely local, remote, and co-rendering, respectively, and provide model compression approaches to facilitate the transmission of radiance field models for rendering. We also present rendering acceleration approaches and joint computation and communication designs to enhance the rendering efficiency. In a case study, we propose a new semantic communication enabled 3D content transmission design, in which the radiance field models are exploited as the semantic knowledge base to reduce the communication overhead for distributed inference. In addition, we discuss the utilization of radiance field rendering in wireless applications like radio mapping and radio imaging, in which radiance field models are used to effectively represent complex radio environments to facilitate wireless network designs. It is our hope that this paper can provide new insights on the interesting wireless integration with radiance field rendering for future 6G networks with 3D contents.
\end{abstract}

\begin{IEEEkeywords}
6G, immersive communications, 3D Gaussian splatting (3D-GS), neural radiance field (NeRF), federated learning, inference.
\end{IEEEkeywords}

\maketitle

\section{Introduction}
Sixth-generation (6G) networks are experiencing a paradigm shift from the connected everything in fifth-generation (5G) to the new vision of connected intelligence, bridging the virtual and physical worlds. On the one hand, 6G networks are envisioned to utilize new wireless technologies such as millimeter wave (mmWave)/terahertz (THz) and extremely large-scale antenna arrays, which can provide ultra-high-data-rate, hyper-reliable, and ultra-low-latency communications to handle the massive data transmission demands. On the other hand, 6G networks are expected to evolve towards new multi-functional networks by integrating various functionalities such as wireless sensing, communication, mobile edge computation (MEC), and artificial intelligence (AI) \cite{itur2022, Guangxu2023push, Liu9737357ISAC}. The multi-functional operation facilitates the fusion of emerging AI technologies and wireless networks, and thus helps meet the diverse requirements of emerging applications such as extended reality (XR), metaverse, autonomous driving, and intelligent robotics.

In this paper, we focus on a particular usage scenario of 6G, namely immersive communications, which aim at creating a highly engaging and interactive environment for communication users, with typical applications including XR, telepresence, immersive gaming, and metaverse \cite{Shen_2023}. Different from the enhanced mobile broadband (eMBB) applications in 5G that focus on the transmission of two-dimensional (2D) contents, the realization of the new immersive communications applications in 6G highly relies on the transmission and processing of three-dimensional (3D) contents. In particular, the 3D contents can represent the spatial environment and object information, accurately depict real physical scenes, and effectively model real physical worlds for, e.g., telepresence and immersive gaming. Due to the rich spatial information contained, 3D contents can also be used for more efficient object detection and localization in 3D environments, facilitating the navigation and decision-making of auto-driving vehicles and intelligent robots. Therefore, to fully exploit the benefits, how to efficiently represent, transmit, and reconstruct 3D contents over wireless networks is becoming an important research topic in the 6G era, which requires new interdisciplinary design approaches, by employing techniques from various areas including wireless communications and networks, computer vision, computer graphics, and AI.

Proper representation of 3D contents is the foundation for their efficient transmission and reconstruction over 6G networks. Different from 2D contents that are normally represented via pixels, there are various different approaches for 3D representation. Conventionally, 3D representation approaches can be categorized into explicit and implicit ones \cite{ahmed2019survey, li2024advances}. In general, explicit representations are characterized by well-defined formats such as mesh \cite{Bommes12014mesh}, voxel \cite{s21248241}, and point cloud \cite{5980567} to intuitively present the shape and surface of 3D objects to viewers, while implicit representations typically employ implicit functions to express contents in an implicit manner such as occupancy functions \cite{1011452732197}. For instance, point cloud \cite{5980567} is a widely adopted explicit 3D representation approach, which represents 3D objects by organizing numerous data points to depict the geometry without complex data processing. These discrete points, however, fall short in depicting the details of 3D objects such as textures. Different from conventional representation approaches, neural radiance field (NeRF) \cite{nerf3503250} and 3D Gaussian splatting (3D-GS) \cite{{Kerbl3DGS}} have recently emerged as new transformative 3D representation approaches based on radiance field rendering. These approaches represent the distribution of light and depict the interactions between light and surfaces, materials, and surroundings, thus effectively representing the details of 3D contents and rendering photorealistic images from novel views. 

NeRF and 3D-GS have their respective pros and cons. On the one hand, NeRF innovatively represents 3D contents by employing neural networks, which have exceptional capabilities in visual reconstruction. In particular, NeRF only needs to utilize multi-view images of a scene as training data to generate novel viewing images via volume rendering, which is able to represent the detail of 3D contents in high visual quality with only lightweight models \cite{gao2023nerf}. Nevertheless, NeRF involves dense sampling of light rays and volume rendering, which require substantial computation resources and are time-consuming in general. On the other hand, 3D-GS \cite{Kerbl3DGS} is a promising explicit radiance field representation approach, which extends the point-based rendering by utilizing Gaussian functions to effectively represent the scene, thus avoiding dense ray sampling. As compared to NeRF, 3D-GS explicitly represents 3D contents with a well-structured data format, which can be edited easily and is able to show the content geometry and properties more intuitively. Furthermore, the training process of 3D-GS only requires to optimize Gaussian functions and the rendering process allows for parallel computation without querying neural networks and dense ray sampling, which are generally more computationally efficient. However, the explicit representation of 3D-GS requires a large amount of data and may pose challenges for storage and transmission, especially for the representation of large-scale scenes. As a result, NeRF models are lightweight but computationally heavy, while 3D-GS is efficient in computation but less efficient in storage and transmission. 

Due to the rapid advancements of NeRF and 3D-GS for 3D representation in the computer society, it is envisioned that these radiance field rendering approaches will play an important role in immersive communications applications over 6G networks. Therefore, it is becoming necessary and urgent to embrace the radiance field rendering in 6G networks. However, such integration is still in its infancy, which introduces various new technical challenges to be tackled. In particular, the radiance field models (especially the explicit 3D-GS models) are represented by a large amount of data and require excessive resources for storage and transmission, and the training and inference (or rendering) of radiance field models (especially the implicit NeRF models) require intensive computation. While conventional radiance field training and rendering are implemented in a centralized manner at remote cloud or at local devices, 6G wireless (edge) networks have distributed communication, computation, and storage resources at networked base stations (BSs) and mobile devices. Therefore, it is demanding to investigate new distributed training and inference methods for the training and rendering of radiance fields by taking into account the practical constraints on communication, computation, and storage resources at distributed nodes. Along this direction, new joint computation and communication designs are crucial. Furthermore, immersive communications applications such as telepresence normally have stringent end-to-end latency requirements on the transmission, reconstruction, and display of  3D contents to ensure the quality of experience (QoE) \cite{Shen_2023}. This thus calls for a unified design of the whole sensing-communication-computation processing pipeline over the 6G wireless networks to minimize the end-to-end latency while preserving the QoE requirements. In addition, due to the complicated wireless environment, user mobility, and the broadcast nature of wireless signals, the transmission of radiance field models for distributed training and inference may experience significantly fluctuated wireless channels over time and space, and face severe multi-path channel fading and co-channel interference. The unreliable and dynamically changing wireless communication channels thus make the distributed training and inference more difficult. As such, new training-and-inference-task-oriented wireless design approaches need to be devised jointly with the radiance field optimization.

To address the above challenges, this paper provides a comprehensive investigation on the integration of the two representative radiance field rendering methods (i.e., NeRF and 3D-GS) in 6G wireless networks. Specifically, in Section \ref{overview} we present an overview on the basics of radiance field rendering, NeRF, and 3D-GS, and accordingly point out their potential applications and implementation challenges over wireless networks. In Section \ref{NeRFtrain}, we discuss the over-the-air training for NeRF and 3D-GS models. In particular, we present the federated learning design over hierarchical device-edge-cloud architectures, in which various important issues such as
joint computation and communication resouces management and camera (device) placement are discussed in detail. We also discuss potential extensions by considering the issues such as synchronization, generalization, vertical federated learning, and over-the-air federated learning. In Section \ref{NeRFrendering}, we discuss rendering architectures for the over-the-air rendering or inference of radiance fields by utilizing distributed computation and communication resources. We discuss the model compression to improve the transmission efficiency, and present acceleration approaches and joint computation and communication designs to enhance the rendering efficiency. More specifically, we propose a novel semantic communication enabled inference design by using the radiance field models as the semantic knowledge base to reduce the communication overhead for distributed inference, which is validated via a case study of 3D human face transmission in Section \ref{casestudy}. In Section \ref{NeRFempower}, we show that radiance field models can also be employed to benefit wireless network design in return, in which radiance field models are employed for radio mapping, radar imaging, and multi-modal-sensing-assisted communications. Finally, Section \ref{conclusion} concludes this paper.

\section{Radiance Field Rendering Over Wireless Networks} \label{overview}

\begin{figure*}
	\centering
        \captionsetup[subfloat]{font=footnotesize}
	\subfloat[NeRF]{
		\begin{minipage}{15cm} 
                        \includegraphics[width=\textwidth]{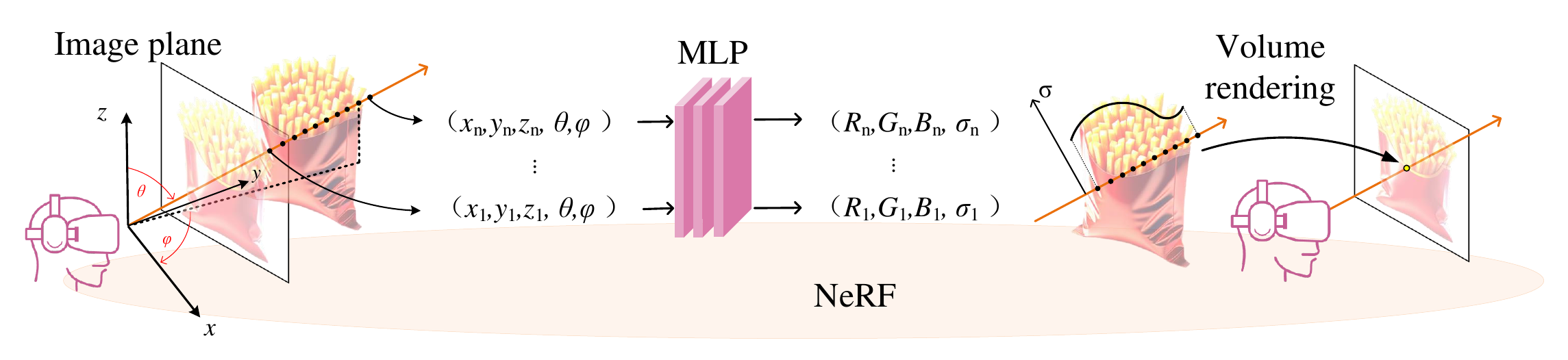} \\
		\end{minipage}
	}
        \captionsetup[subfloat]{font=footnotesize}
\\
	\subfloat[3D-GS]{
		\begin{minipage}{15cm}
			\includegraphics[width=\textwidth]{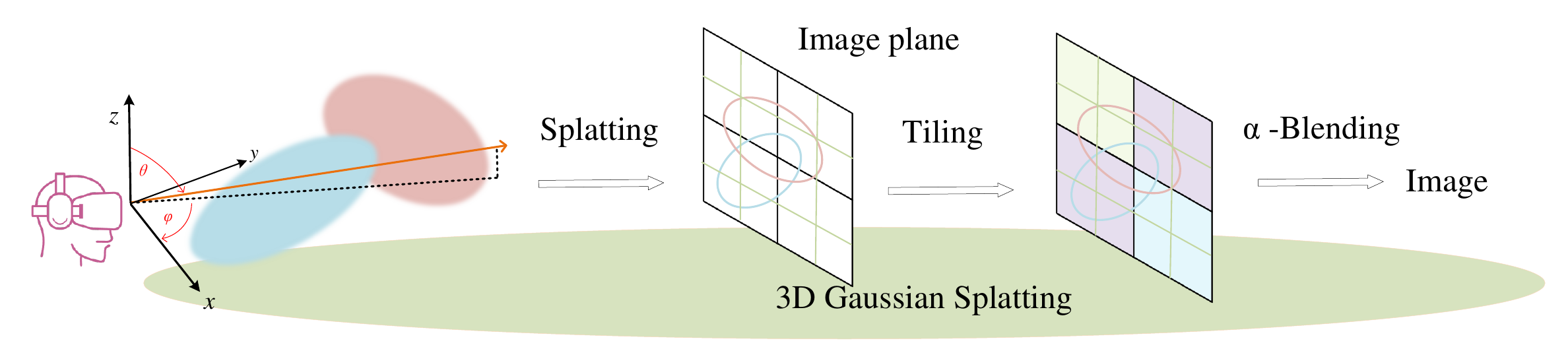} \\
			
		\end{minipage}
	}
\caption{Basics of NeRF and 3D-GS based on radiance field rendering.} 
\label{NeRFand3DGS}
\end{figure*}

This section introduces the exploitation of radiance field rendering over wireless networks to support applications with 3D contents. First, we review the basics of radiance fields and introduce NeRF and 3D-GS. Then, we discuss potential 3D-content-based applications by using NeRF and 3D-GS in wireless networks. Finally, we present new technical challenges faced by the integration of NeRF/3D-GS over wireless networks.  
\subsection{Basics of Radiance Field Rendering}
The concept of radiance fields originates from the physical properties of light. Based on how light interacts with surfaces, materials, and surroundings along its propagation path, we can create 3D spatial radiance fields with distinct light distribution. Radiance fields can be used to describe scenes in 3D space via a five-dimensional function $f_{\text{radiance}}$, which maps a 3D location $\hat{\boldsymbol{x}}=(x,y,z)$ and a viewing direction in sphere $\boldsymbol{d}=(\theta, \phi)$ to the RGB color space $\boldsymbol{c}=\left(R,G,B\right)$ and the opacity/volume density value $\sigma$, i.e.,
\begin{align}
    f_{\text{radiance}}(x,y,z,\theta, \phi)=\left(R,G,B,\sigma\right).
\end{align}
In general, the representations of radiance fields can be divided into explicit and implicit ones, respectively. The implicit radiance field representation uses neural networks to represent radiance information, based on which the color information and opacity/volume density are derived by querying the neural networks. In contrast, the explicit radiance fields intuitively present the radiance information with discrete spatial data, in which each spatial data element stores color information and opacity. Among various design approaches, NeRF \cite{nerf3503250} and 3D-GS \cite{Kerbl3DGS} are becoming the most prominent implicit and explicit radiance field representations, respectively, which are detailed in the following.

\subsubsection{NeRF} \label{NeRFoverview}

As shown in Fig.~\ref{NeRFand3DGS}(a), NeRF utilizes multilayer perceptron (MLP) neural networks to approximate the radiance information in 3D space, which implicitly characterizes the RGB color and volume density of every volume point within the 3D scene. In particular, the NeRF model takes the coordinates of volume point $\hat{\boldsymbol{x}}=\left(x,y,z\right)$ and viewing direction unit vector $\boldsymbol{d}=(\theta,\phi)$ as input, and outputs the corresponding RGB color $\boldsymbol{c}=\left(R,G,B\right)$ and volume density $\sigma$ of every volume point. These volume points are virtually generated by mapping the pixels to camera rays along the given viewing direction $\boldsymbol{d}$. After querying the MLP, the colors and densities of volume points are accumulated based on volume rendering, thereby determining the pixels of 2D images from given viewing directions one by one. It is worth noting that the process of volume rendering is differentiable and thus the MLP network can be trained by using the mean square error (MSE) between the sampled images and the rendered results as the loss function. As the MLP networks implicitly store the color and density information of each sampled volume point, lightweight models are sufficient for NeRF-based representation, which only require a relatively small amount of data. However, due to the implicit representation, NeRF rendering requires frequent querying of neural networks to obtain the color and density information for different volume points, and also needs complex integral calculations during the volume rendering process, thus resulting in high computation loads.

\subsubsection{3D-GS}

As shown in Fig.~\ref{NeRFand3DGS}(b), 3D-GS is an explicit representation approach that utilizes a set of Gaussian functions to represent the 3D scenes. In 3D-GS, each Gaussian $\mathcal{G}$ contains coordinate parameter $\hat{\boldsymbol{x}}=(x,y,z)$, 3D covariance $\Sigma \in \mathbb{R}^{3 \times 3}$, opacity $\alpha$, and color $c$, i.e., $\mathcal{G}=\{\hat{\boldsymbol{x}},\Sigma,\alpha,c \}$. Here, the color and opacity information are explicitly stored in the Gaussian function $\mathcal{G}$, without relying on the querying of neural networks. As such, to synthesize an image from a specified viewing direction $(\theta, \phi)$, we first need to project the set of Gaussians into the 2D imaging space determined by the viewing direction, in which each pixel in the 2D space can be covered by multiple projected Gaussians. Next, we segment the image plane into tiles and identify the projected Gaussians that cover each tile. Finally, we perform rendering, in which the color of each tile is calculated through alpha blending. As the rendering process of 3D-GS is differentiable, backpropagation based on gradient descent methods can be utilized to optimize the parameters of Gaussian functions for taining 3D-GS models, similarly as for NeRF. The benefits of 3D-GS are the improved rendering efficiency and the reduced computation overheads. In particular, the explicit representation of 3D-GS allows for easy access to the color and opacity information without relying on querying neural networks, and each independent tile allows for parallel rendering. However, the use of explicit color and opacity data in 3D-GS requires a large amount of data for representation, introducing new challenges in storage and communication overloads.

In summary, the implicit representation of NeRF is able to represent 3D contents by a lightweight model and render realistic images without deteriorating details of 3D contents, but it requires the frequent querying of neural networks for volume rendering and thus is computationally inefficient. By contrast, the explicit representation of 3D-GS is intuitive and easy to edit and can render high-quality images without relying on neural networks or volume rendering and enables parallel rendering processes, but it requires a large amount of data for representation, transmission, and storage. Therefore, there generally exists a tradeoff between the communication, computation, and storage in selecting the 3D representation approaches when implementing radiance field rendering in diver scenarios over wireless networks. 

\subsection{Applications in Wireless Networks}

\begin{figure*}
    \centering
    \includegraphics[width=1\linewidth]{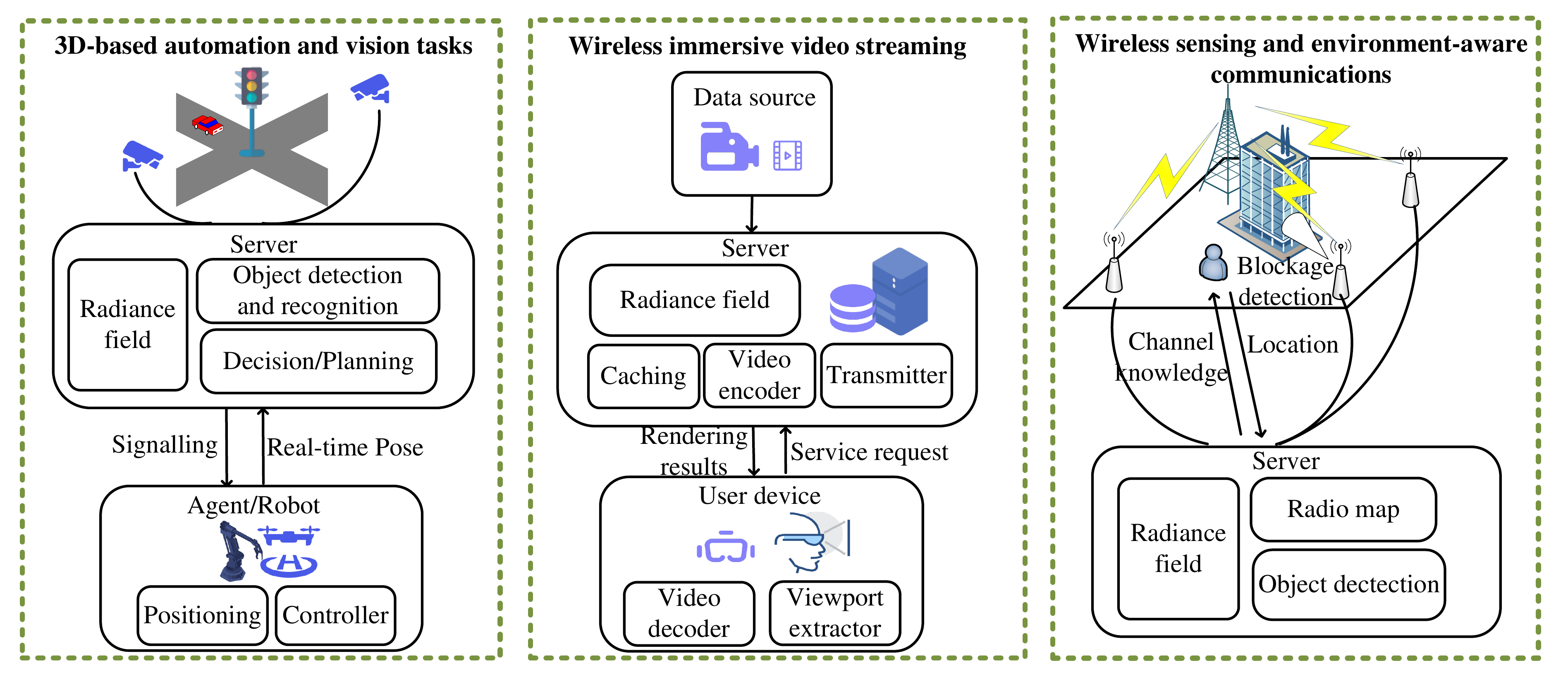}
    \caption{Typical wireless applications of radiance field rendering. }
    \label{Typicalapplication}
\end{figure*}

NeRF and 3D-GS have made great impacts in the areas of computer vision and computer graphics. Their excellent capability to represent 3D contents has also attracted growing interests from the communications society for supporting various 3D-content-enabled applications. This subsection provides three typical applications of NeRF and 3D-GS in 6G networks, as shown in Fig.~\ref{Typicalapplication}.
\subsubsection{\textbf{Wireless immersive video streaming}}
3D video technologies such as $360^\circ$ video \cite{9133103} and volumetric video \cite{jin2023capture} have emerged as new paradigms in the multimedia field, which are crucial to enable immersive communications. Unlike conventional 2D video, $360^\circ$ video and volumetric video can provide immersive visual experiences. In particular, $360^\circ$ video can provide users with three degrees of freedom (3DoF) view (i.e., pitch, yaw, and roll), and volumetric video explores more immersive virtual scenes with 6DoF (including pitch, yaw, roll, and additional x-, y-, and z-axis). Recently, the emergence of NeRF and 3D-GS are expected to find promising applications for wireless 3D video streaming, providing immersive vision experiences to large-scale mobile users. This is due to the fact that NeRF and 3D-GS can outperform traditional point cloud and mesh representations in rendering photorealistic video frames \cite{jin2023capture,chen2024survey}. In particular, 3D-GS is advantageous in real-time radiance field rendering, which makes it particularly suitable for wireless 3D video streaming when the computation resources are limited at devices. By contrast, NeRF models are lightweight in general and can be transmitted efficiently with less communication overhead, which makes it particularly useful for the scenario when the communication rates among different nodes are constrained \cite{Jedari9252131VideoSurvey}. One prominent application of NeRF and 3D-GS is telepresence \cite{jin2023capture, ChengTelepresence}, which captures and transmits a more realistic representation of the human body, shape, and facial expressions, thereby significantly enhancing the immersive communication experience. Other applications include XR, immersive gaming, and digital avatars, in which the content provider can represent the 3D content with high-quality NeRF or 3D-GS models, and then transmit these models to the edge networks (e.g., remote edge servers or devices) for rendering and display 3D videos to a large number of end users.

\subsubsection{\textbf{3D-based automation and vision tasks}}
Radiance field rendering also holds significant potential to enhance 3D-based vision intelligence tasks such as video analytics \cite{Xu_2023_ICCV}, simultaneous localization and mapping (SLAM) \cite{tosi2024nerfs}, and robot navigation \cite{Adamkiewicz9712211Navigation}. This is primarily attributed to the role of spatial information provided by 3D contents, which significantly enhances the inference capabilities required for these tasks. For instance, NeRF/3D-GS models can reconstruct high-quality scenes containing terrain information to facilitate video object detection and robot navigation. In the 6G era, these tasks are implemented at wirelessly connected agent/robots and remote servers in a distributed manner \cite{wang2024mulanwc, 2024distributed}, by fully leveraging their distributed computation and communication resources. For example, in a robot navigation task, the robots can transmit their captured multi-view images to the edge server for training NeRF/3D-GS models, based on which, the edge server can subsequently render images of novel viewing regions by leveraging the rich computation resource. Next, the server can transmit these novel-viewing images back to the robots to assist their global path planning and decision making. In these applications, the training and inference of radiance fields at the network edge need to be designed by properly exploiting the distributed computation and communication resources. 

\subsubsection{\textbf{Wireless sensing and environment-aware communications}}
Motivated by the great success in the field of optical imagery, radiance field rendering is expected to be applicable in the wireless field, especially for the representation of wirelessly generated 3D information by effectively modeling the physical process of wireless signal propagation. This application is becoming increasingly important for 6G networks, as radio sensing and imaging are seamlessly integrated as a new function of wireless networks with the advancement of integrated sensing and communication (ISAC) \cite{Liu9737357ISAC, Lyu9916163, Lu10418473}. Radiance field rendering can benefit the fusion of sensing and communication functionalities by enhancing the sensing performance with 3D spatial information and provides multi-modal information to facilitate the communication designs. In particular, radiance field rendering such as NeRF can be used to model the radio frequency (RF) signal attenuation process including signal reflection, scattering, and refraction, and accordingly represent radio maps that store the channel information at different transceiver locations \cite{zhao2023nerf}. In addition, radiance field rendering is also useful for radar imaging (e.g., via synthetic aperture radar (SAR)) \cite{lei2023sarnerf}, in which the radio imaging process can be modeled by using NeRF, such that the SAR imaging information can be obtained from different views. Furthermore, radiance field rendering can benefit sensing-assisted environment-aware communications \cite{Zeng10430216radiomap}, in which radiance field models can be used to represent the 3D environment with considerable visual quality and multi-view vision information, such that the environmental obstacles and scatterers can be detected to infer and predict the channel knowledge information \cite{Charan9512383Proactive, Feifei10412143Vision}.

\subsection{Technical Challenges}
Due to the effectiveness in 3D content representation, NeRF and 3D-GS have emerged as promising techniques for the efficient transmission and reconstruction of 3D contents over wireless networks. Nevertheless, the integration of these radiance field rendering approaches in 6G networks introduces several new technical challenges that need to be addressed.

\subsubsection{\textbf{Intensive computation demands in radiance field training and rendering}} The training and rendering of radiance fields are generally computationally intensive. On the one hand, the training of neural networks and volume rendering processes in NeRF are computationally heavy and time-consuming. For example, the vanilla NeRF model \cite{nerf3503250} needs a convergence time of 1-2 days when training on a single graphics processing unit (GPU), which is impractical for many real-time applications. On the other hand, though not relying on neural networks querying or volume rendering, 3D-GS still needs to deal with a large number of Gaussian functions that can be computationally intensive, especially in the case with large-scale scenes. In the literature, there have been various works \cite{Hedman_2021_ICCV, Alexander3528223ngp, Wu_2022_CVPR} focusing on improving the training and rendering efficiency for NeRF and 3D-GS from the algorithmic perspectives, which, however, do not consider their implementation over wireless networks. Different from conventional centralized cloud, wireless edge networks consist of a large number of distributed BSs and mobile devices with heterogeneous and highly distributed computation resources. As such, it is essential but challenging to design distributed training and rendering approaches for radiance fields over such heterogeneous wireless networks.

\subsubsection{\textbf{Communication overheads for transmission of radiance fields}}

The distributed training, distributed inference, and deployment of radiance fields at wireless edge networks need frequent exchange of NeRF and 3D-GS models across different nodes. This may lead to significant communication overheads. First, from the training perspective, substantial observation views are required to construct a fine radiance field. This massive data delivery introduces great challenges in bandwidth-constrained wireless networks. Next, from the model transmission perspective, radiance field models representing large-scale scenes may require a large model size for transmission, especially for 3D-GS that utilizes millions of Gaussians \cite{Kerbl3DGS}. Therefore, it is important to explore the model compression methods for reducing the overhead of transmission. In addition, considering executing inference over the air, only a portion of the entire 3D scene associated with the viewport of user is needed for visualization, as the user normally only sees a portion of a 3D object/scene from a specific viewing direction. In this case, we need to develop task-specific designs to only render interested regions for reducing the communication overhead while preserving the QoE, instead of rendering the full scene that may incur substantial communication overhead. Last but not least, wireless transmission experiences significantly fluctuating wireless channels and faces channel multi-path fading and interference, which may significantly degrade the communication data rate. In this case, the rendering process needs to be devised jointly with the wireless transmission by considering the bandwidth limitations and wireless channel fading. 

\subsubsection{\textbf{Critical end-to-end latency requirements}}

Intelligent applications like telepresence and XR have stringent requirements on end-to-end information processing latency for real-time transmission and rendering \cite{Shen_2023}. In practice, the radiance field models can be deployed at cloud center, end devices, edge servers, or distributed at both end users and edge servers. In the rendering process, end users or viewers need to determine the viewing directions, such that the rendering task can be performed to obtain the images under the viewing directions, which are then  transmitted  and displayed to the viewers. The above process involves both computation and communication in general, which jointly determine the end-to-end rendering latency. Depending on the deployment locations of the radiance field models, there generally exists a trade-off between the communication and computation latency. For instance, the communication latency becomes significant when the radiance field models are deployed at far-apart cloud with huge computation power, and the computation latency is essential when the models are deployed at local devices with limited computation power. By contrast, deploying these models at edge servers of BSs or deploying them at both edge servers and end devices may achieve balanced communication and computation latencies, as the edge servers have high computation power and are close to end users or viewers. To optimize the efficiency of radiance field rendering in these scenarios, it is important but challenging to pursue joint computation and communication designs.

\section{Training of Radiance Field Models over Wireless Networks} \label{NeRFtrain}
This section considers the training of radiance field models over wireless networks, in which the NeRF or 3D-GS models are trained to represent the 3D scenes, by using pre-stored or real-time collected dataset with multi-view images at a number of distributed mobile devices. First, we briefly introduce two general training architectures, namely centralized learning and distributed (federated) learning, respectively. Next, we consider a particular hierarchical device-edge-cloud federated radiance field learning architecture to support the training of radiance field models for large-scale scenes. Furthermore, we discuss critical design issues for federated radiance field learning. Finally, we discuss some important extensions along this direction.

\subsection{Centralized versus Distributed Learning}
The training of radiance field models can be implemented via centralized and distributed learning in general, which are detailed in the following, respectively.
\subsubsection{Centralized Radiance Field Learning} In centralized radiance filed 
learning, there exists a centralized server (e.g., a cloud center or a powerful edge server) that collects the multi-view dataset from mobile devices over the wireless network, and then trains the NeRF or 3D-GS models via using the collected dataset by leveraging its rich computation capabilities. This centralized learning is suitable for scenarios when the computation resources are rich at the server but severely limited at end devices. This may correspond to a scenario with distributed urban-block internet-of-things (IoT) cameras with limited computation power, which can collect multi-view image data and then send them back to the monitoring center for training the radiance field models of the concerned block. However, this leads to increased communication overhead due to the massive raw data transmission, and also introduces privacy issue for distributed devices. 

\subsubsection{Distributed (Federated) Radiance Field Learning} In distributed radiance field learning, a number of separate nodes use their distributed computation resources to collaboratively train radiance field models in a distributed manner, by keeping the multi-view data distributed instead of sharing them to a central node. Among various approaches, federated learning \cite{mcmahan17a} is particularly appealing for distributed radiance field learning. The federated radiance field learning is normally coordinated by a central node that can be the central cloud or an edge server at BS, and is implemented in an iterative manner. In each iteration, participating devices independently train their local models by using their own observed multi-view data and only need to transmit model parameters to the central node. After collecting the radiance field model parameters from devices, the central node aggregates them to obtain an updated global model and then distributes it back to the participating devices. The above iterations will terminate until convergence. The federated radiance field learning design utilizes distributed computation resources at participating devices without sharing the local multi-view raw data, thus avoiding the data leakage and preserving the data privacy. Due to the frequent NeRF/3D-GS model exchange between participating devices and the central node, the communication overheads are becoming the performance bottleneck.

In summary, the centralized radiance field learning is able to process the multi-view data and train the radiance field model in a central network node without coordinating distributed devices, but it may incur large communication overheads and privacy leakage issues. By contrast, the federated radiance field learning aims at exploiting the distributed computation resources among separate devices to train the models without sharing raw data, which, however, generally requires large iterations of model exchange between devices and servers.

\subsection{Hierarchical Federated Radiance Field Learning}
\begin{figure*}
    \centering
    \includegraphics[width=0.65\linewidth]{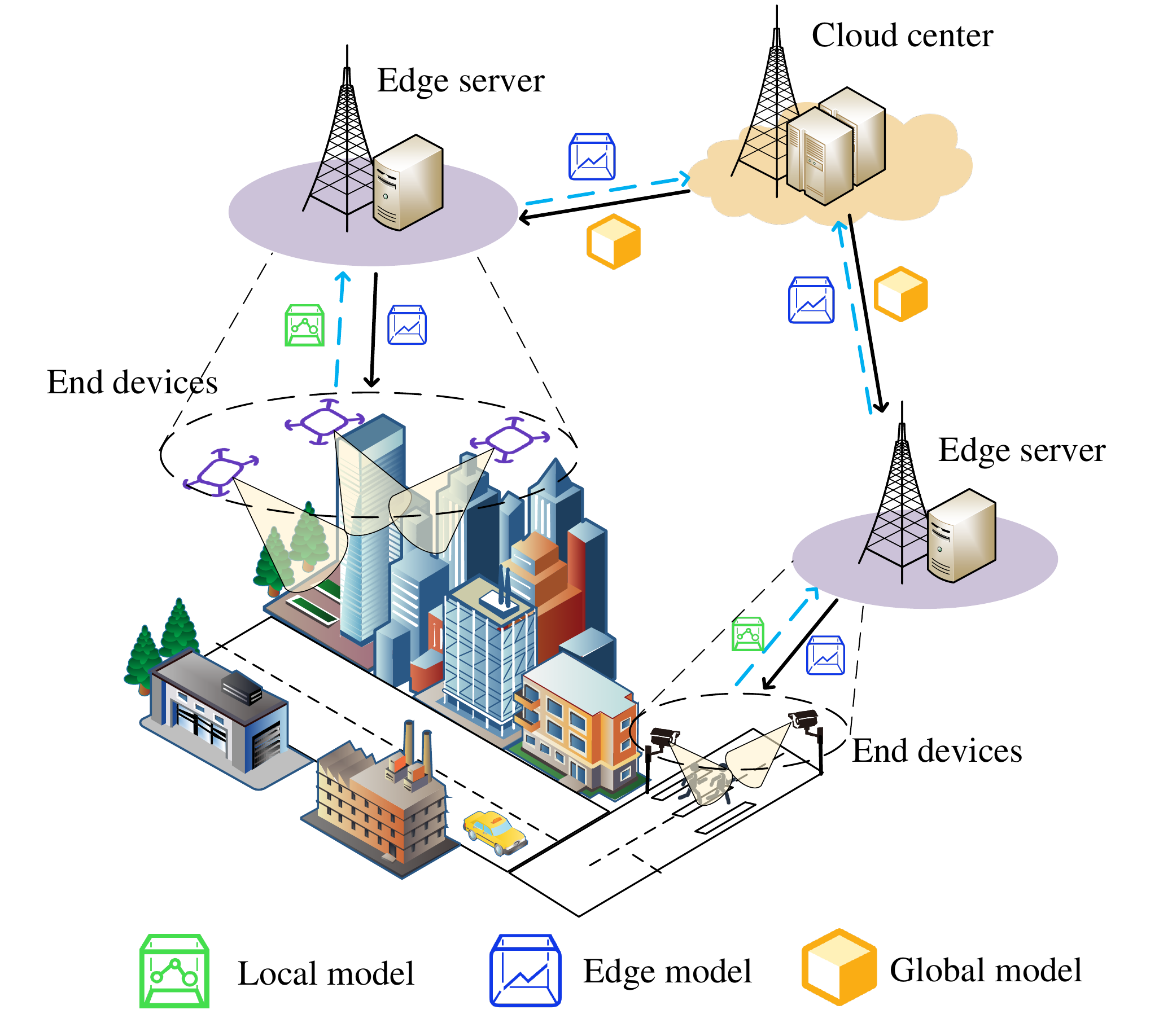}
    \caption{Hierarchical federated radiance field learning over 6G networks with end devices, edge servers, and cloud center.}
    \label{NeRFTraining}
\end{figure*}
Due to the benefits in exploiting distributed computation power and preserving data privacy, federated radiance field learning is particularly appealing for training over wireless networks. Nevertheless, prior works on federated radiance field learning focused on designing the learning algorithm, without considering specific wireless network architectures. In this paper, we present a new device-edge-cloud hierarchical federated learning architecture for the training of radiance field models for a large-scale scene.\footnote{ Generating 3D contents for large-scale scenes may have abundant applications in various 6G applications like low-altitude economy \cite{cheng2024networked}, smart city \cite{Tancik_2022_CVPR}, and intelligent transportation \cite{2024distributed}. The design principles for large-scale scenes are also applicable to the case with small-scale scenes \cite{holden2023federated}.}  Specifically, this architecture consists of one cloud server, $L>1$ edge servers, and a large number of $K > L$ end devices, as shown in Fig.~\ref{NeRFTraining}. At the lower level of the hierarchical architecture, the end devices are each installed with one or more cameras to capture the multi-view observation data, and the edge servers each cover a certain area to coordinate multiple associated end devices. At the higher level, the cloud center coordinates the training operation of these edge servers. The device-edge-cloud hierarchical federated radiance field learning approach is appealing for the efficient training of large-scale scene radiance fields (e.g., at a city level). In particular, the hierarchical architecture allows the cloud center to coordinate massive end devices through a limited number of edge servers, thus significantly reducing the traffic loads over the whole network. Next, each end devices only need to send their local model information to nearby edge servers for aggregating edge models of a relatively smaller-scale sub-scene, thus enhancing the training efficiency. Furthermore, the cloud center can aggregate the edge models from edge servers to update the global models for the large-scale scene. In addition, the cloud center can also adjust the learning parameters (e.g., frequencies of edge and global aggregations) and coordinate the network-level resource management,  by fully exploiting the diverse communication, computation, and storage powers over the whole network.

To begin with, we briefly introduce the dataset for training the radiance field models. Let $D_k=\{(I_{k,i},p_{k,i})\}$ denote the local dataset at device $k\in \{1,\ldots,K \}$, where $I_{k, i}\in \mathbb{R}^{H \times W \times 3}$ denotes the $i$-th observed image by device $k$ that serves as the ground truth during the training process. Here, $H$ and $W$ denote the height and width of the observed image, respectively. Furthermore, $p_{k,i}$ denotes the corresponding viewing information, which contains the extrinsic matrix and intrinsic matrix of the cameras. The extrinsic matrix exploits the rotation and translation information to transform world coordinates into camera coordinates, while its inverse matrix can convert camera coordinates back to world coordinates. The intrinsic matrix uses the focal information to project camera coordinates onto the 2D image plane. In computer vision society, there are two methods to obtain the dataset. The first method is to directly extract images and the corresponding viewing information from synthetic datasets \cite{nerf3503250}. The second method is to utilize the COLMAP structure-from-motion package \cite{Schonberger_2016_CVPR} to estimate the extrinsic and intrinsic matrices for real observed images \cite{Mildenhall3306346}.

Then, the hierarchical federated radiance field learning is implemented as follows by particularly considering the use of 3D-GS models.  Due to the inherent limitations in camera views, each device can only access the multi-view observation data from a subsection of the entire scene. Under this setup, the training of 3D-GS models is implemented over nested two rounds of iterations. Specifically, we denote the local 3D-GS model at device $k$ as $G_{\text{local}}^k = \{ \mathcal{G}^{(i)}\} = \{\hat{\boldsymbol{x}}^{(i)},\Sigma^{(i)},\alpha^{(i)},c^{(i)} \}_{i}$, which is a set of 3D Gaussians. Similarly, we denote the edge 3D-GS model at edge server $l$ as $G_{\text{edge}}^l = \{\hat{\boldsymbol{x}}^{(i)},\Sigma^{(i)},\alpha^{(i)},c^{(i)} \}_{i}$ and the global model at cloud center as $G_{\text{global}}$. In each lower-level or inner iteration, the end devices each perform local training (via multiple rounds of stochastic gradient descent (SGD) calculations) based on the captured data and then upload their local models and the 3D location information to their associated edge servers. After receiving the local models and local location information, each edge server aligns the pose of its associated devices in a common coordinate, and then aggregates the received local models to obtain updated edge models. In particular, each edge server $l$ obtain the edge model via merging the local models as $ G_{\text{edge}}^l = G_{\text{local}}^1 \cup G_{\text{local}}^2 \cup ,\ldots, \cup~G_{\text{local}}^{\hat{k}}$, where $\hat{k}$ denotes the number of associated end devices of edge server $l$. Then, in each upper-level or outer iteration, the edge servers transmit the updated edge models (after one or more rounds of inner iterations) and the corresponding coordinate information to the cloud center, such that the cloud center can align the coordinates and then aggregate the edge models to obtain the global models. Similarly, the cloud center obtains the global model via merging the edge models as $G_{\text{global}} = G_{\text{edge}}^1 \cup G_{\text{edge}}^2 \cup,\ldots, \cup~G_{\text{edge}}^{L}$. In general, the frequencies of inner and outer iterations (i.e., the number of local SGD calculations within each inner iterations and the number of inner iterations or edge aggregations within each outer iteration) may affect the consumed computation and communication resources towards convergence, and thus are key decision variables for performance optimization. 

Notice that the pose alignment presents a unique challenge in federated radiance field learning for large-scale scenarios, where end devices (such as UAVs) are dispersed across various locations, each observing the scene from unique viewing directions and thus obtaining the multi-view dataset with different pose information $p_{k,i}$ \cite{suzuki2024federated, suzuki2024fed3dgs}. In this case, the local models obtained at different end devices have distinct poses, which cannot be directly aggregated at the edge server. As such, the edge server needs to first use proper pose alignment to revise these poses towards the common coordinate, and then perform the edge aggregation. As such, it is critical for the end devices (and edge servers) to upload their pose information (and coordinate information) to facilitate the edge (and cloud) aggregation. Some useful device pose alignment methods can be referred to in, e.g., \cite{suzuki2024federated,suzuki2024fed3dgs}.

\subsection{Joint Resource Management and Device (Camera) Placement}
For the hierarchical federated radiance field learning over wireless networks, it is important to pursue joint computation and communication resource management to improve the convergence speed for training. Towards this end, it is desirable to first analyze the convergence behavior for training radiance field models. Various prior works provided analytic results on the convergence behaviors of general federated edge learning systems in terms of optimality gap \cite{Cao10422876aircomp}, averaged gradient norm bound \cite{Bottou}, and generalization gap \cite{NIPS2017_a5e0ff62}, as a function of computation and communication resources as well as the federated learning parameters. These results may be tailored for the federated radiance field learning of our interest. Next, based on the convergence results, we can perform joint resource management to enhance the convergence performance under constrained resources, by jointly optimizing communication (e.g., transmit power and bandwidth) and computation (e.g., computation frequency and learning rounds) resources over the network together with the federated learning parameters (like the number of SGD rounds within each inner iteration and the number of inner iteration rounds within each outer iteration). There are some very interesting tradeoffs between communication and computation that can be exploited for performance optimization. For instance, allowing more SGD rounds in each inner iteration may lead to increased computation at end devices but less lower-level communication between the end devices and edge server (due to the fast convergence with less rounds of iterations needed). Similarly, employing more inner iteration rounds may increase the computation and communication overheads in the low-level network, but may reduce those in the upper level. There are many interesting problems worth future investigation.

Another important factor affecting the learning performance is the camera/device placement. On the one hand, placing cameras at proper locations with proper poses helps acquire high-quality multi-view raw data   \cite{2022ActiveNeRF}, thus reducing the amount of data required for training radiance field models and the associated computation overheads. On the other hand, due to the randomness of wireless environment, a good placement location for observing multi-view data may not have good channel conditions for wireless transmission, and thus may lead to increased communication latency. Therefore, optimizing the camera placement locations is an interesting direction to enhance the training efficiency by properly balancing the underlying tradeoff between computation (due to data quality) versus communication (due to channel quality). In the literature, ActiveNeRF \cite{2022ActiveNeRF} is an efficient method to construct high-quality multi-view training data under limited observation views. Specifically, ActiveNeRF introduces an uncertainty parameter into the original NeRF architecture as the additional output to identify and select novel views that can most significantly benefit model performance. It will be interesting to combine  ActiveNeRF for camera deployment with joint resource management to further enhance the training performance over wireless networks.

\subsection{Extensions}
Besides the above federated radiance field learning design, there are also some interesting extensions by considering the asynchronous model aggregation,  generalization issue, vertical federated learning, and novel over-the-air computation (AirComp) for highly-efficient aggregation. We discuss these extensions to motivate future work. 
\subsubsection{Asynchronous Federated Radiance Field Learning} 
The above federated radiance field learning design requires synchronous learning among different end devices and edge servers, which, however, can be challenging due to the heterogeneity of different nodes. From the perspective of end devices, the local training process varies due to the heterogeneity of their computation capabilities, and thus it is difficult to synchronize the local model training and aggregation across diverse end devices. From the perspective of data, local multi-view data have different distributions across different end devices, thus frequent uploading of models on particular end devices may incur divergence to the global radiance field model, resulting in overfitting to specific datasets (i.e., specific observation views). A promising solution is to execute the asynchronous learning, in which the aggregation and construction of radiance field models are conducted upon receiving local models from a subset of devices, rather than postponing it until the entire local models have been uploaded from all devices \cite{xie2020asynchronous}. In such an asynchronous learning paradigm, proper device selection becomes important. For instance, we need to
evaluate the contributions of different devices on the learning of radiance field models over wireless environments, such that those devices that significantly contribute to the overall model convergence have higher priorities to be activated in the learning process \cite{Yu3582377Async, XU2023100595Asynchronous}. 

\subsubsection{Generalizable Federated Radiance Field Learning} 

Generalization is another crucial issue for evaluating 3D representations.  First, limited by inadequate training views, NeRF and 3D-GS overfit to these sparse training views, resulting in poor visualization performance from novel views. Next, a well-trained radiance field model may be scene-specific and struggle to represent different or dynamic scenes. In the computer vision community, researchers spend much effort to enhance the generalization ability \cite{Chen_2021_ICCV, Yu_2021_CVPR}. One promising solution is to first train a global model based on training data from different scenes and then fine-tune this model into a personalized one based on the private local data \cite{Chen_2021_ICCV}. When these pre-training and fine-tuning processes of radiance field models are implemented distributedly at edge servers and end devices, it is important to explore the computation and communication resources management designs to balance the training accuracy and latency \cite{lyu2024rethinking}.

\subsubsection{Vertical Federated Radiance Field Learning}
Vertical federated radiance field learning is another interesting issue, where datasets from different devices share the same sample space but differ in feature spaces \cite{QiangYang10415268VFL}. This is the case when  depth cameras are practically involved in generating depth maps (reflecting spatial information of scenes), in addition to the conventional cameras generating multi-view RGB images. To leverage the overall data collected from RGB and depth cameras for efficient radiance field model learning, we should consider vertical federated learning paradigm together with depth images based radiance field construction algorithms (see, e.g., \cite{Deng_2022_CVPR}).

\subsubsection{Over-the-air Radiance Field Federated Learning} In radiance field federated learning, a large number of devices from different views may participate 
for high-quality and large-scale 3D model training, where communication delay becomes the performance bottleneck due to the frequent exchange of model parameters. To alleviate such challenges, AirComp \cite{Zhu9535447Aircomp} has emerged as a promising solution to enable the so-called over-the-air learning (see, e.g., \cite{Cao10422876aircomp}), where devices concurrently transmit their radiance field models/gradients for simultaneous aggregation at the server. By exploiting the superposition property of wireless multiple-access channels, AirComp could dramatically improve the communication efficiency for aggregating radiance field models, but may introduce a new type of AirComp errors that may degrade the training convergence \cite{Cao10422876aircomp}. How to invoke AirComp into federated radiance field learning, and accordingly perform learning performance analysis (especially by taking into account the effect of AirComp errors) and overall system resource management are interesting directions. The synchronization of hierarchical and heterogeneous devices in over-the-air radiance field federated learning is a critical problem, especially for the training of large-scale radiance field models.

\subsubsection{Radiance Field Federated Learning for Dynamic Scenes} Dynamic scene representation is an important issue in radiance field rendering, which has been investigated in prior works \cite{Pumarola_2021_CVPR, Yang_2024_CVPR, Wang_2022_CVPR}. In general, there are two methods to represent the dynamic scenes. Firstly, for NeRF, deformable neural networks are widely adopted to represent dynamic scenes. For instance, the work \cite{Pumarola_2021_CVPR} embeds an additional time dimension into the model to control the color and density for time-varying scenes. This allows the model to account for dynamic changes over time. For 3D-GS, deformable fields are exploited to dynamically control the position and 3D covariance of Gaussians over time \cite{Yang_2024_CVPR}. This approach enables the 3D-GS model to adjust dynamically as the scene changes, providing accurate representations of dynamic scenes. Inspired by these methods, we can employ federated learning to train deformable radiance field models to represent dynamic scenes. Furthermore, another promising method for dynamic scene representation is to first train a generalized radiance field model to predict coarse color and density information and then fine-tune it to predict finer details for accurately rendering dynamic scenes \cite{Wang_2022_CVPR}.

\section{Deployment and Rendering of Radiance Field over Wireless Networks} \label{NeRFrendering}
Once the radiance field models are well-trained, we need to deploy them over wireless networks for efficient rendering. In this section, we first discuss three basic architectures for radiance field rendering. Next, we discuss the techniques of model compression and acceleration for enhancing the efficiency of model transmission and rendering. Furthermore, we provide various joint computation and communication designs for radiance field rendering.

\subsection{Architectures for Radiance Field Rendering }
\begin{figure*}
    \centering
    \includegraphics[width=0.7\linewidth]{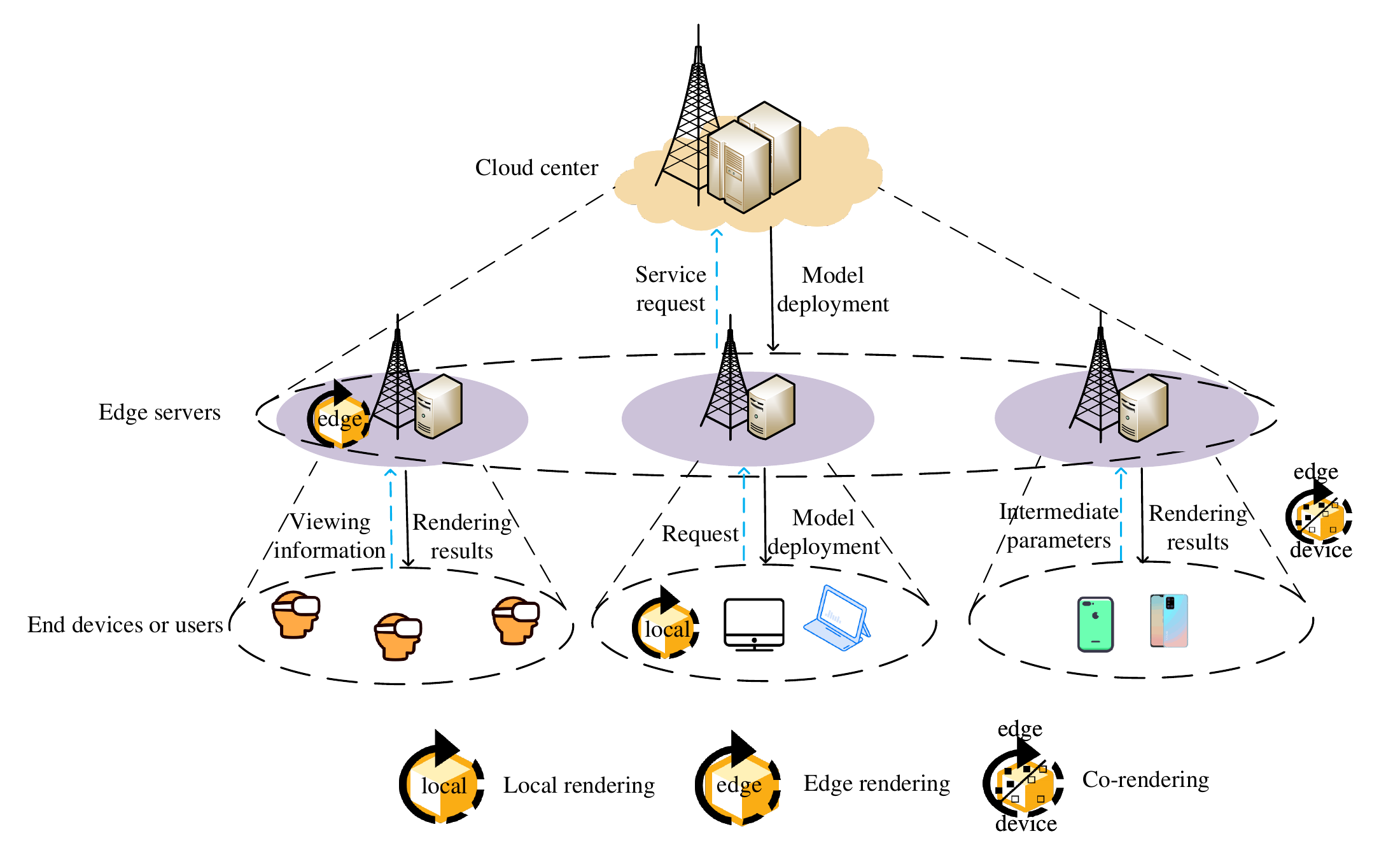}
    \caption{Deployment and rendering of radiance fields over 6G networks.}
    \label{NeRFRendering}
\end{figure*}

Based on the locations deploying radiance field models, the architectures for rendering are generally categorized as three types, namely, edge rendering, local rendering, and co-rendering, respectively,  as illustrated in Fig.~\ref{NeRFRendering}. 

\subsubsection{Local Rendering}
For local rendering, radiance field models are deployed at end devices such as computers, mobile devices, and virtual reality (VR) headsets, and accordingly the rendering process is conducted locally at these devices. To implement local rendering, the trained radiance field models are transmitted to be deployed at local devices, such that local devices can execute rendering instantly by using its local computation power, and  display the content based on the viewing information of end users. The local rendering can generally reduce the rendering latency due to the local processing, which, however, may only work well when the radiance models are lightweight, due to the limited computation and storage capabilities at end devices.

\subsubsection{Edge Rendering}
For edge rendering, radiance field models are deployed at edge servers close to end devices. In the edge rendering process, the end devices first send viewing information to edge servers, and then the servers can render images on the viewing information and send back the rendering results to the end devices. By leveraging the cloud-like computation and storage capabilities at edge servers, edge rendering can support the rendering of large radiance field models in a swift manner. Nevertheless, the two-way round-trip communication between the edge servers and end devices becomes an important issue limiting the rendering efficiency.

Another design issue for enhancing the efficiency of edge rendering is to select proper edge servers for model deployment and accordingly determine the server-device associations. On the one hand, the deployment of radiance field models needs to not only consider the channel conditions between the end devices and the associated edge servers, but also the traffic load balancing among multiple device-server pairs.  On the other hand, the end devices may move randomly over time, and the associated edge servers need to change dynamically, thus further complicating the deployment and migration issues. 
The edge server selection and server-device association need to comprehensively consider issues like edge traffic conditions, wireless channel conditions, as well as user mobility for efficient edge rendering.

\subsubsection{Co-Rendering}
To fully utilize the computation resources at edges and devices as well as exploit both benefits of local and edge rendering, device-edge co-inference or co-rendering has emerged as a promising solution, which has been widely adopted in various AI applications nowadays \cite{Angela9758628Split}. In the device-edge co-inference or local-remote co-rendering, the radiance field models (e.g., the MLP neural network for NeRF) are split into two parts to be deployed at end devices and edge servers, respectively, such that the rendering is implemented distributedly at the two types of nodes. In the co-rendering process, the end devices first obtain the opacities and colors of sampled points along the rays based on the deployed partial radiance field models (e.g., MLPs). These colors and opacities serve as the intermediate parameters, which are then sent to the server for volume rendering. Finally, the server sends the rendering results back to the devices. The co-rendering design is expected to enhance the rendering performance via properly designing the task allocation between the devices and servers, for which the joint communication and computation designs become essential.  

In summary, the local rendering, edge rendering, and co-rendering have their pros and cons in terms of rendering latency and computation/storage resource requirements. First, local rendering deploys trained radiance field models on local devices, enabling instant rendering by leveraging local computational resources and avoiding long-distance transmission. This approach minimizes latency but demands lightweight models due to limited local computational and storage capacities. Next, edge rendering allows the end devices to send viewing information to edge servers, which render images and then send the rendering results back. This may result in substantial latency due to two-way communication, but can handle larger models due to the richer computational resources at edge servers. Furthermore, co-rendering combines the strengths of local and edge rendering by utilizing distributed resources of end devices while offloading heavy tasks to edge servers, thus improving the overall rendering performance. Despite its efficiency, co-rendering may still encounter latency issues from the intermediate parameter exchange between devices and servers. Therefore, the choice of proper rendering architecture should consider specific latency requirements and available computational/storage resources.

\subsection{Model Compression and Acceleration}
The compression of radiance field models and the algorithmic rendering acceleration designs are important to enhance the efficiency for the transmission/deployment and rendering of radiance fields.\footnote{Note that network routing and load balancing are also important for the transmission of radiance field models over large-scale communications networks when there are a large number of edge servers and end devices demanding various different radiance field models. These issues, however, are out of the scope of this paper on wireless edge networks in 6G, which are interesting topics for future work.}

\subsubsection{Model Compression}
Compressing the radiance field models is essential for model deployment, especially when the training nodes generating the models are far-apart from the end users and the size of trained models becomes large, e.g., for representing large-scale scenes. The model compression can reduce the model size, ensuring faster and more reliable transmission, which is particularly important in scenarios with limited bandwidth or under stringent latency requirements. Efficient radiance field model compression approaches generally include network pruning, quantization, low-rank approximation, and knowledge distillation \cite{jin2023capture, chen2024survey}. For NeRF,  network pruning and weight quantization are efficient to remove the redundant network layers and transform the full-precision floating-point numbers into lower-bit representations, respectively. Furthermore, low-rank approximation is applicable to replace the high-rank matrices with low-rank ones, while knowledge distillation extracts knowledge from a large model and condenses it into a smaller one. Different from NeRF, the size of 3D-GS models depends on the volume of explicit data, i.e., the number of Gaussian functions. To efficiently compress 3D-GS models, pruning \cite{fan2024lightgaussian} and quantization \cite{navaneet2023compact3d} are promising to reduce the number of Gaussian functions for decreasing the amount of explicit data, while maintaining the rendering quality. 

\subsubsection{Algorithmic Model Acceleration} 
The acceleration of rendering from an algorithmic perspective is also useful to reduce rendering latency. This is especially crucial for NeRF, where synthesizing new views based on a well-trained model requires frequent queries to MLPs for generating density and color at each sampling point. This may incur significant time costs in the rendering stage, and thus influence the overall latency of inference. There are generally two approaches from the computer vision community \cite{Hedman_2021_ICCV, Alexander3528223ngp, Wu_2022_CVPR}. One approach is to pre-compute and store the neural networks \cite{Hedman_2021_ICCV}, transforming the original NeRF model into more easily accessible data structures to significantly reduce the rendering time. The other approach is to learn the scene features by additionally utilizing other representations like voxel grids \cite{Wu_2022_CVPR}, allowing for reducing the size of neural networks and leading to less rendering time.

\subsection{Joint Computation and Communication Design}
Edge rendering and co-rendering require intensive computation at edge servers and end devices as well as frequent information transmission between them. Therefore, it is becoming necessary to jointly design the computation and communication for enhancing the efficiency. In the following, we first discuss joint computation and communication resource management, and then introduce new semantic communications for facilitating radiance field rendering.

\subsubsection{Joint Resource Management}
Joint communication, computation, and storage resources optimization has been investigated in the MEC literature for enhancing the service performance. For example, the authors in  \cite{Du9120235ImmersiveVR} considered the immersive VR video transmission design with MEC in THz communication scenarios, where the rendering task offloading and transmission power allocation are jointly designed to minimize total energy consumption. The authors in \cite{Sun8728029CCCVR} proposed a joint communication, caching, and computing design to optimize the mobile VR delivery strategy, providing profound insights into the communication-storage-computing tradeoff. The authors in \cite{Xu10437783CollaborativeRendering} proposed an edge-device collaborative rendering VR framework, in which the rendering tasks are split to be executed at edge servers and end devices by taking into account the practical constraints on bandwidth resources and latency. While these joint resource management designs are promising, they have not considered the specific properties of radiance field rendering for 3D contents. With NeRF and 3D-GS based representations, we can design the rendering resolutions as a new design degree of freedom for NeRF or 3D-GS, together with the joint resource management, for enhancing rendering efficiency. For  3D-GS, we can also design the number of Gaussian functions together with the resource management optimization.   

 \subsubsection{Semantic Communications Enabled Radiance Field Rendering}
Under the local-remote co-rendering architecture, both the server and the device collaboratively execute rendering tasks, where frequent data exchanges may incur significant communication overhead. 
To tackle this issue, semantic communications, a novel communication system design paradigm beyond the conventional bit-level transmission, is becoming an interesting new solution that can be applied to enhance the end-to-end rendering performance. The effective transmission and rendering of 3D contents with semantic communications rely on the semantic knowledge base construction and semantic feature extraction. Specifically, semantic knowledge base provides rich knowledge for semantic information processing (i.e., feature extraction and data recovery) \cite{ren2023knowledge}, which can be exploited as the side information at both the transmitter and the receiver to reduce the transmitted bits while preserving the QoE. In particular, radiance field models can be exploited as high-quality semantic knowledge base storing fruitful radiance information of 3D scenes \cite{guanlin}. With such knowledge base, the transmitter only needs to extract and transmit essential semantic features to the receiver for real-time 3D reconstruction. Furthermore, to accommodate the dynamic channel conditions especially in the wireless environment, channel condition-aware semantic feature selection and transmission policies should be further designed. The semantic communications-based co-inference is appealing for not only reconstructing 3D contents, but also for further performing downstream intelligent tasks such as 3D video analysis. A case study on using semantic communications for 3D human face transmission is presented in the next section to show the value of semantic communications in this direction.

\section{Case Study: Semantic Communication for 3D Content Transmission} \label{casestudy}
\begin{figure*}
    \centering
    \includegraphics[width=0.75\linewidth]{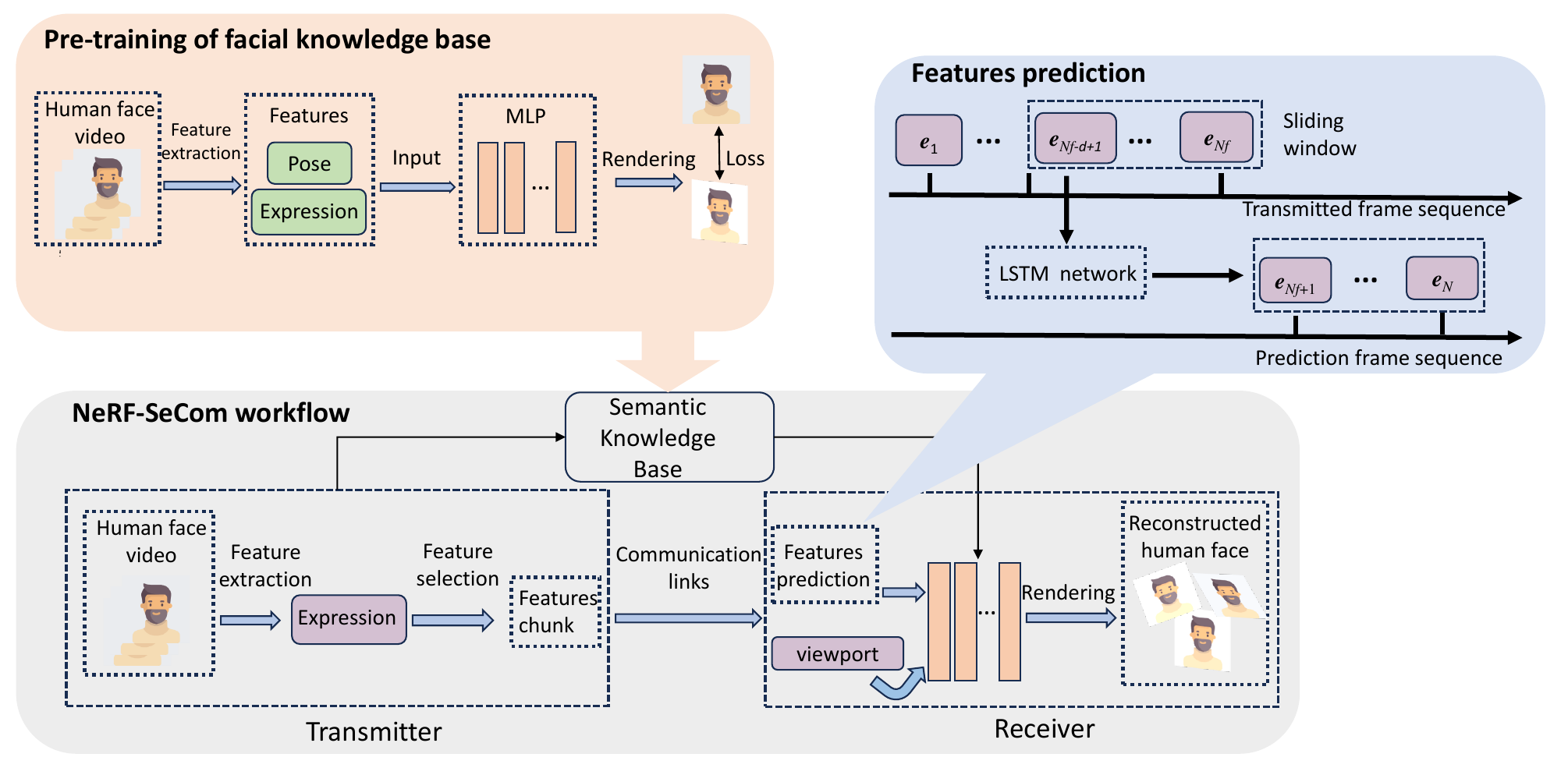}
    \caption{Illustration of proposed semantic communications framework for 3D human face transmission with NeRF.}
    \label{semanticframework}
\end{figure*}
\begin{figure}
    \centering
    \includegraphics[width=0.9\linewidth]{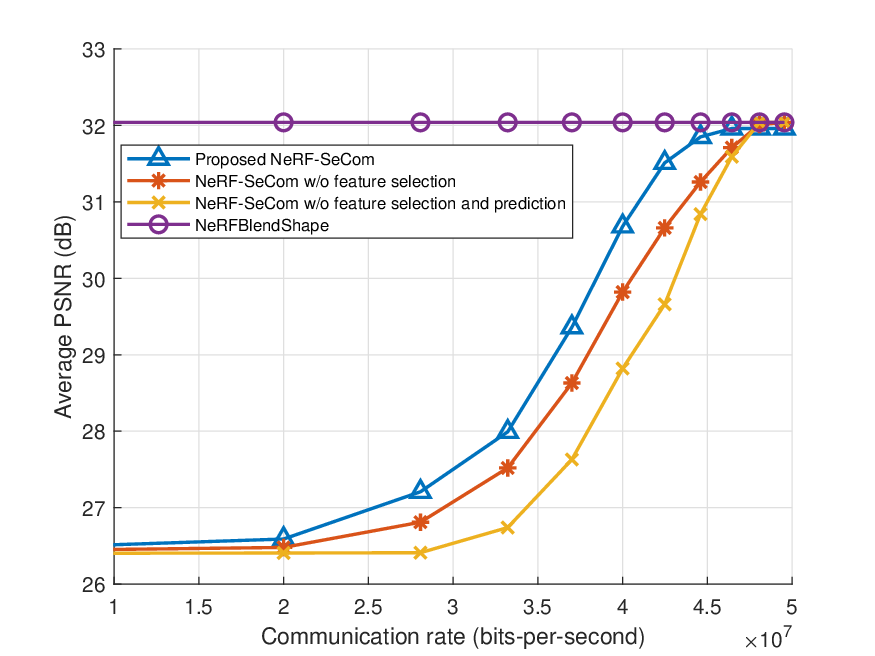}
    \caption{Performance comparison between our proposed NeRF-SeCom framework versus benchmark schemes in terms of average PSNR under variational communication rates.}
    \label{PSNR}
\end{figure}
\begin{figure}
    \centering
    \includegraphics[width=1\linewidth]{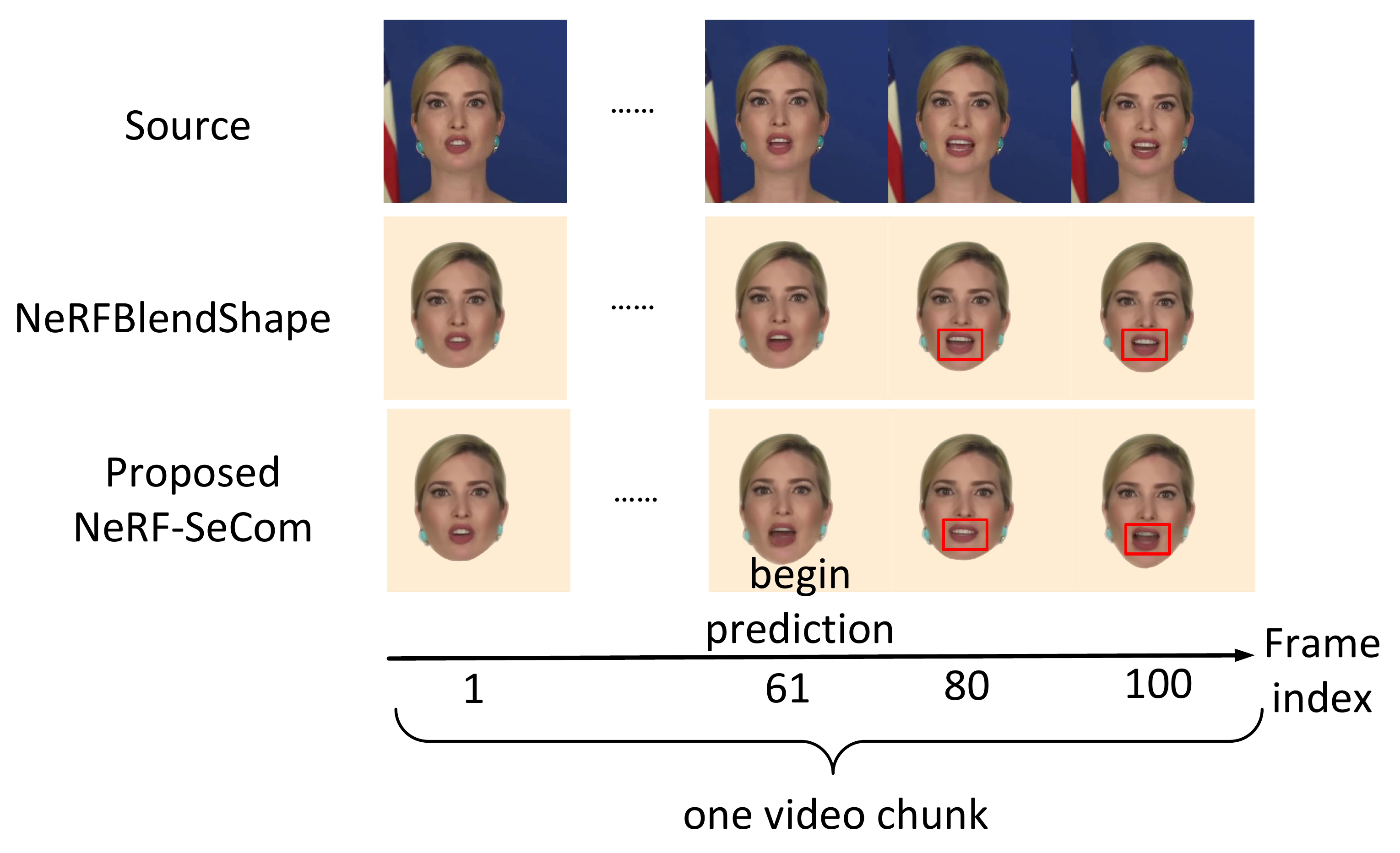}
    \caption{Rendering results comparison sampled from frame $61$, $80$, and $100$ with NeRFBlendShape and ground truth by setting $N_f=60$.}
    \label{semantic}
\end{figure}
This section presents a novel semantic communications framework for the transmission and reconstruction of a particular type of 3D content, i.e., 3D human face, by considering the presentation via NeRF \cite{guanlin}.  In the literature, there have been extensive prior works investigating the semantic transmission of 2D content (see, e.g., \cite{Lyu10388062Semantic, Qin9763856textSecom,ren2023knowledge}). However, there are only a few studies exploiting 3D content transmission via semantic communications \cite{wang2024taskoriented, ChengTelepresence}. The work \cite{ChengTelepresence} proposed to use keypoints, 2D images, and text as the semantics to transmit the 3D content by using the mesh representation. The work \cite{wang2024taskoriented} explored the 3D point cloud wireless transmission by using the keypoints of human avatars as the semantic features. Despite these advancements, 3D content semantic communication with radiance fields, especially NeRF and 3D-GS, has not been studied yet. 

The semantic communications framework with NeRF (NeRF-SeCom) for 3D human face transmission is shown in Fig.~\ref{semanticframework}, which includes face semantic feature extraction, feature selection, feature chunk packing, learning-based feature prediction, and face rendering. Semantic knowledge base aims to provide rich knowledge for semantic information processing (i.e., feature extraction and data recovery), which is essential in semantic communication systems \cite{ren2023knowledge}. We exploit the NeRFBlendShape \cite{Gao3555501nerfface} face NeRF model, which employs a multi-level hash table \cite{Alexander3528223ngp} to associate expression coefficients for accelerating the rendering process. This model serves as the semantic knowledge base, assisting in both face semantic feature extraction and face rendering. The NeRFBlendShape face model is expressed as 
\begin{align} \label{NeRFBlendShapemodel}
    \hat{\mathcal{R}}_\beta\left(C, \boldsymbol{e}\right)\rightarrow (\sigma, \boldsymbol{c}),
\end{align}
where $\hat{\mathcal{R}}$ represents the querying of hash table and MLP networks, $\beta$ represents the learnable weight parameters of the MLP network, and $C$ and $\boldsymbol{e}=\left\{e_1,e_2,\ldots,e_M \right\}\in \mathbb{R}^{M}$ denote the camera parameters and the expression feature coefficients, respectively. Here, $M$ denotes the total dimension of the expression features. NeRFBlendShape adopts the facial expression as the semantic feature to represent 3D faces. Accordingly, the human face based on NeRFBlendShape is rendered by
\begin{align}
    \hat{\boldsymbol{I}}=V(\mathcal{R}_\beta\left(C, \boldsymbol{e}\right)),
\end{align}
where $V(\cdot)$ represents the volume rendering. In our implementation, we first pre-train the NeRFBlendShape-based facial knowledge base with a segment of monocular video of human faces as the training dataset in an offline manner. Next, the pre-trained semantic knowledge base is shared at the transmitter and receiver before 3D face transmission. The semantic knowledge base corresponds to the facial identity,\footnote{The research on generalizable face models applicable to different identities is indeed an interesting direction and holds significant potential. How to generate such models is beyond the scope of the current work and can be considered for future research.} and as a result only needs to be trained and transmitted once for the same person over a long time period consisting of a large number of video chunks. It is worth noting that the NeRFBlendShape model is lightweight and thus it can be transmitted without influencing the communication latency in real time. In the following, we provide the detailed workflow of our proposed NeRF-SeCom framework. 

First, semantic feature extraction is performed on the input video at the transmitter. The face feature extraction follows from the facial blendshape method \cite{Cao6654137FaceWarehouse}, in which each expression coefficient has a corresponding specified semantic meaning. Then we obtain the expression coefficients and the face pose parameters as $\boldsymbol{e}=\left\{e_1,\ldots,e_M\right\}$ of each frame, where $M$ represents the dimension of expression coefficients.

Next, features are packed into a video chunk. Unlike traditional video streaming, which encodes a batch of video frames into video chunks for each transmission, our framework packs the extracted features into chunks. To efficiently reduce the communication overhead for transmission, we classify the expression features into static and dynamic types. Accordingly, for each chunk, we allow the transmitter to send the average value of static expression coefficients only once for the whole video chunk with $N$ frames, and transmit the dynamic expression coefficients for a number of $N_f \leq N$ frames to adhere to the rate constraint. In particular, suppose that the total number of dynamic expression features is $M_{\text{dyn}}$, and each expression coefficient in one frame is quantized to a size of $Q$ bits. Therefore, the transmitted bits in the first frame are $Q M$, and the transmitted bits in the remaining frames are $Q(N_f-1) M_{\text{dyn}}$. Then we have the total bits to be transmitted as $Q M + Q(N_f-1) M_{\text{dyn}}$, which should not exceed $\tau R$, with $\tau$, $R$ denoting the latency and communication rate, respectively. Accordingly, we have $Q M+ Q(N_f-1) M_{\text{dyn}} \leq \tau R$.

After receiving the expression features in the whole chunk, the receiver renders the 3D human face by utilizing the NeRF models in the shared facial knowledge base. Here, the users can freely change their viewing directions to render the viewpoints of the 3D content. However, the face rendering needs the complete  expression features over all frames, but only dynamic expressions in the first $N_f$ frames are transmitted. Therefore, we need to develop proper feature prediction methods for estimating the dynamic expressions in the remaining $N - N_f$ frames based on the received features. In particular, we develop a long short-term memory (LSTM) network for prediction. Specifically, we utilize the expression parameters from previous frames in the sequence to predict the expression parameters of subsequent frames. On the receiver side, the received expression features $\left\{\boldsymbol{e}_1,\boldsymbol{e}_2,\ldots,\boldsymbol{e}_{N_f}\right\}$ of each chunk are fed into the LSTM network to generate the predicted expression coefficients $\left\{\boldsymbol{e}_{N_f+1},\boldsymbol{e}_{N_f+2},\ldots,\boldsymbol{e}_{N}\right\}$.

To demonstrate the effectiveness of our proposed framework, we conduct the following experiments under the communication rate constraints. First, we compare the quality of facial reconstruction of our proposed NeRF-SeCom framework with the benchmarks that deactivate feature selection scheme or deactivate both feature selection and prediction schemes. We evaluate the performance via the average peak signal-to-noise ratio (PSNR) in one chunk (i.e., 100 frames). Second, we compare the visual rendering results on different frames by our proposed framework, versus those by NeRFBlendShape (without communication), and the source images.

Fig.~\ref{PSNR} shows the performance of our proposed NeRF-SeCom framework and the benchmark schemes under variational communication rate. It is observed from Fig.~\ref{PSNR} that our proposed framework maintains satisfied reconstruction performance under different data rates, and the performance increases as the communication rate increases. Moreover, in high rate regime, the performance of our proposed framework converges to the upper bound since the communication rate can ensure the full transmission of total video frames in one chunk. Next, it is also observed that the proposed framework outperforms the benchmark schemes significantly, which demonstrates the effectiveness of our proposed framework with feature selection and prediction. However, in high and low rate regimes, the performance gaps between the proposed framework and benchmark schemes are limited. This is due to the fact that, on the one hand, in low rate regimes, long time-step predictions lead to accumulated errors. On the other hand, in high rate regimes, the need for feature selection and prediction significantly reduces due to the sufficiently high communication rate for feature transmission.

Fig.~\ref{semantic} shows the comparison of visual rendering results among our proposed framework, NeRFBlendShape, and the source images under $N_f=60$. We provide the rendering results of the 61-st frame (the first frame to start prediction), the 80-th, and the 100-th frame (the last frame) for comparison. It is observed that our proposed framework can render high-quality facial images compared to the NeRFBlendShape scheme. It is also observed that as the amount of predicted frames increases, the error of the rendered face rises. This is due to the fact that each shift of the sliding window incorporates the current predicted results as true values for subsequent predictions. In such a case, the accumulation of prediction errors from the previous sequence deteriorates the system performance.

It is worth emphasizing that although this case study focuses on 3D human face transmission by exploiting the NeRF face model as the semantic knowledge base, our proposed semantic framework can be well generalized to other 3D contents. First, this framework can be implemented in the 3D human avatar transmission by exploiting avatar radiance field models such as \cite{NEURIPS2021Su} as the semantic knowledge base to facilitate semantic information processing. The effectiveness of this paradigm has been demonstrated by the prior works that exploit conventional 3D representations as the semantic knowledge base in human avatar transmission. For example, the work \cite{wang2024taskoriented} proposed a point cloud semantic communication framework, in which an avatar model storing skeleton and appearance information is exploited as the knowledge base to help semantic information extraction and avatar pose recovery, but this work did not consider radiance field model. Besides, our framework can be implemented in the transmission of complex scenes. We can exploit a radiance field model to store the features of scenes like \cite{Chen_2021_ICCV} as the semantic knowledge base, which is a generic model pre-trained on large scene datasets. By leveraging this model at both transmitter and receiver, we only need to transmit features extracted from the images of target scenes at the transmitter and fine-tune the pre-trained scene model based on the received features at the receiver. Accordingly,  we can reconstruct the scene following the volume rendering. This is an interesting direction worth investigating in future work.

\section{Radiance Field Rendering for Wireless Applications} \label{NeRFempower}
The previous sections focused on how to use distributed computation and communication resources to support the training and rendering of radiance fields over 6G networks. Motivated by the success of radiance field in representing the distribution of light, it is envisioned that radiance field rendering is promising for the representation of wirelessly generated 3D information based on the physical process modeling of wireless transmission. This is becoming increasingly vital for the development of 6G networks with sensing integration \cite{Liu9737357ISAC} and environment awareness \cite{Zeng10430216radiomap}. In this section, we discuss three wireless applications using radiance field rendering. First, we discuss the utilization of radiance field rendering to facilitate radio map construction and radar imaging (particularly SAR imagery), respectively. Finally, we introduce the use of radiance field rendering for multi-modal-sensing-assisted communications.

\subsection{Radio Map}

Radio map is a location-specific database that stores the wireless environment information, providing channel knowledge based on the locations of transmitter and receiver \cite{Zeng10430216radiomap}. By using the knowledge of RF signal propagation in radio map, we can predict the wireless environment to enable efficient environment-aware communications and flexible frequency management. Motivated by the fact that neural networks in NeRF are effective in modeling light propagation, the work \cite{zhao2023nerf} proposed to use neural networks to model the RF signal propagation, thus constructing an RF radiance field. More specifically, the RF radiance field takes the positions of transmit antennas and receive antennas and the signal measuring direction as input, and outputs the received signals and the attenuation from the measuring direction. As such, the RF radiance field performs as an implicit radio map. Given the position of the transmitter, we can predict the received signals at specific positions. Compared to the conventional approaches, the radiance field-based radio map achieves better performance due to the excellent capability of neural networks to simulate the extremely complex wireless environment. It is interesting to exploit radiance field-based radio map to enable various designs in wireless networks, such as transceiver beamforming, channel estimation, and wireless infrastructure deployment.

\subsection{Radar Imaging}
Radiance field rendering also holds potential to enhance the performance of radar imaging. Radar imaging, especially SAR, has achieved great success in the field of remote sensing due to its robust ability to sense targets and generate images under poor light and weather conditions. However, SAR imaging suffers from its sensitivity to viewing angles. This property makes it difficult to exploit conventional deep learning techniques to effectively learn the features of SAR imagery and infer images from novel viewing angles. To address this issue, the work \cite{lei2023sarnerf} proposed to use neural networks to learn the distribution of attenuation coefficients and scattering intensities, and then render images based on the mapping and projection algorithms. This method demonstrates superior performance over conventional techniques in generating high-quality images from multiple viewing angles. Such advancements can be utilized to improve the recognition and detection capabilities of applications in remote sensing.

\subsection{Multi-modal-sensing-assisted Communications}

Radiance field rendering can be exploited to obtain rich environment information from multi-modal data to help predict the wireless channel information and construct the channel knowledge map. First, NeRF and 3D-GS have been successfully used to effectively represent light distribution in 3D space to render multi-view vision images and depth images, which can be utilized to improve the performance of detecting objects in 3D space by using spatial information. Accordingly, the detected objects are useful to predict the environmental blockages and scatterers for inferring the channel state information, and accordingly construct the channel knowledge maps. Next, radiance fields are also used to model the wireless signal propagation and render the RF signals, which can be exploited to assist BS deployment, wireless sensing, beamforming, and wireless resource allocations. There are some existing works to explore the utilization of vision information to assist communication \cite{Charan9512383Proactive, Feifei10412143Vision}. However, these works use 2D single-view images, without considering spatial information. Therefore, exploring multi-modal-sensing-assisted communications with radiance field rendering is a promising research direction and requires further investigation.

\section{Conclusion} \label{conclusion}
This paper provided a comprehensive overview on the innovative integration of radiance field rendering in 6G networks, by particularly focusing on the over-the-air training and inference of NeRF and 3D-GS for support various emerging intelligent 6G applications. First, we provided a brief review on radiance field rendering techniques, highlighting their applications and technical challenges encountered when implementing over wireless networks. Then, we discussed the architectures and techniques to train NeRF and 3D-GS models over the air, by paying particular attention to the federated learning design over a hierarchical device-edge-cloud architecture.  Next, we discuss the practical rendering architectures of NeRF and 3D-GS models in wireless networks to enable over-the-air inference by using distributed computation and communication resources. We proposed a new semantic communication enabled inference design, in which radiance field models act as a semantic knowledge base, thereby reducing communication overhead and optimizing rendering efficiency. Finally, we discussed the application of radiance field rendering in wireless fields, such as radio mapping, radar imaging, and multi-modal-sensing-assisted communications, in which NeRF models are used to effectively represent complex radio environments for supporting environment-aware wireless network design. Overall, this paper aims to shed light on the technical feasibilities and advantages of embracing NeRF and 3D-GS in 6G networks and open up new prospects for leveraging 3D contents in enhancing future wireless networks. It is our hope that the insights presented in this paper will inspire further research and development in the field, pushing the boundaries of 6G networks. 

\bibliography{IEEEabrv,reference}

\bibliographystyle{IEEEtran}

\end{document}